# Microscopic triaxial cranking model:
# preliminary predictions for oscillator potential and light nuclei

P. Gulshani[1] and A. Lahbas[2]

[1]3313 Fenwick Crescent, Mississauga, Ontario, Canada L5L 5N1, parviz.gulshani@outlook.com
[2]ESMaR, Faculty of Science, Mohammed V University in Rabat, Morocco, a.lahbas@um5r.ac.ma

Corresponding author: P. Gulshani is the corresponding author

The conventional cranking model for uniaxial and triaxial rotation (CCRM3) is frequently used to study rotational features in deformed nuclei. However, CCRM3 is semi-classical and phenomenological because it uses a constant angular velocity, resulting in a number of Hamiltonian symmetry breakings. To investigate the effect of a dynamic angular velocity, a quantal, microscopic, $D_2$, signature, and time-reversal invariant cranking model for triaxial rotation (MSCRM3) is derived exactly from a unitary rigid-flow-velocity-field rotation transformation of the nuclear Schrodinger equation and Hartree-Fock method. Except for a microscopically-determined angular velocity and normally-small residual terms, the MSCRM3 and CCRM3 Schrodinger equations are identical in form. The microscopic angular velocity emerges from the HF variation, and the residual terms may become significant for triaxial rotation at high spins and in back-bending regions where $\Omega_k$ and $\hat{J}_k$ undergo large changes. A preliminary application of MSCRM3 and CCRM3 to the light nuclei $^{20}$Ne, $^{24}$Mg, and $^{28}$Si using the simple self-consistent deformed harmonic oscillator potential predicts some interesting differences between the rotational features and phenomena predicted by the two models. These differences are attributable to the impact of the dynamic angular velocity. It is important to investigate and understand these differences for the simple and realistic nuclear interactions.

## 1. Introduction

The conventional cranking model for triaxial collective rotation (CCRM3) is described by the Schrodinger equation:

$$\hat{H}_{cr}|\Phi_{cr}\rangle \equiv \left(\hat{H}_o - \vec{\Omega}'\cdot\vec{\hat{J}}\right)|\Phi_{cr}\rangle \equiv E_{cr}|\Phi_{cr}\rangle \quad (1)$$

where $\hat{H}_o$ is the nuclear Hamiltonian, $\vec{\hat{J}}$ is the angular-momentum-operator vector, and the $\vec{\Omega}'$ is a constant angular-velocity vector. The model, with a rotationally non-invariant $\hat{H}_o$ (for example, the Nilsson's model) is frequently used to predict uniaxial-rotation properties of deformed nuclei [1-20]. However, it is recognized [2,3,21] that the model is semi-classical and phenomenological. It uses a constant adjustable angular velocity $\vec{\Omega}'$ parameter. Therefore, Eq. (1) is time-reversal and $D_2$ and signature non-invariant, and ignores the interaction between $\vec{\Omega}$ and $\vec{\hat{J}}$. There have been many revealing investigations [3,6,21-25] to derive the model from first principles using various methods and discover the implicit model assumptions and approximations. The major difficulty in deriving a tractable microscopic quantal cranking model has been the strong coupling between the rotation and intrinsic motions.

In this article, we derive a microscopic, quantal, self-consistent, time-reversal and $D_2$ invariant cranking model for triaxial rotation (MSCRM3) with an angular velocity that is microscopically determined and containing no free parameters.

The article is organized as follows. In Section 2, we derive from a unitary-rotation transformation of the nuclear Schrodinger equation to an intrinsic rotor Schrodinger equation, and then apply Hartree-Fock (HF) method to the rotor equation. The microscopic angular velocity emerges from the application of the HF variation. The intrinsic-rotation coupling in the transformed equation is completely eliminated using a rigid-flow prescription for the rotation angles in the unitary transformation, and requiring the angles to be canonically conjugate to the angular momentum. We note that the choice of the rotation angles (for example, rigid or irrotational, etc flow) is arbitrary and does not change the energy. It changes the term representing the coupling of the intrinsic and rotational motions, and hence the division of the Hamiltonian between these two motions. Except for a microscopically-determined angular velocity and normally-small residual terms, the MSCRM3 and CCRM3 Schrodinger equations are identical in form. The residual terms may become significant for triaxial rotations at high spin and back-bending regions where the angular velocity and angular momentum undergo large changes.

The method used in this article contrasts with that used in [3,6,21-25], which were mostly limited to a rotation in two dimensions and used a significant number of major approximations and assumptions such as large deformation, large expectation of an angular momentum component, expansion in powers of the angular momentum, angular momentum projection, and generator co-ordinates method using various simplifying approximations such as axial symmetry, expansion in density matrices and limiting the analysis to single-particle density matrix, etc.. In particular, reference [23], which uses an approach somewhat similar to that in this article, assumes a rotation in two dimensions and uses an angular-momentum-projected well-deformed independent-particle HF ground-state wavefunction possessing a large value of the expectation of the square of the angular momentum. It obtains a Hamiltonian that depends on the angular momentum $\hat{J}$ and $\hat{J}^2$ operators where the $\hat{J}$ term in the Hamiltonian involves a product of $\hat{J}$ and the individual particle momenta. This $\hat{J}$ term represents the coupling between the rotational and intrinsic motions. Reference [23] then minimizes the expectation of $\hat{H}$ in a particle-hole HF approximation expanded up to second order in $\hat{J}^2$ to obtain an expression for the effective moment of inertia similar to that of Thouless-Valatin.

In Section 3, we use MSCRM3 and CCRM3 with the simple self-consistent deformed harmonic oscillator potential without pairing and spin-orbit interactions to predict rotational features in the light nuclei $^{20}$Ne, $^{24}$Mg, and $^{28}$Si. CCRM3 with this potential was also used in [2-5,7,8,10-19] to predict the nuclear rotational properties. The preliminary and exploratory calculation reported in this article provides information on the impact of the microscopic angular velocity and identifies differences between MSCRM3 and CCRM3 predicted rotational features and yields physical insight. It also provides information on the types of rotations, rotation-shape transitions, and other phenomena that may occur in a planned detailed calculation using realistic nuclear interactions. Therefore, no comparison with measured data is performed in this article. Only for the case of $^{20}$Ne, where pairing interaction is insignificant, we use the spin-orbit and residuals of nuclear interaction and the square of the angular momentum to compare the predicted and measured excitation energies at $J=8$ at which the rotation becomes uniaxial. We report the predicted quadrupole moment for $^{20}$Ne, where we expect and obtain only a qualitative agreement with the measurement because we have not included the spin-orbit interaction in the calculation of the quadrupole moment in the models.

Section 4 summarizes the MSCRM3 derivation, predictions, and the differences between MSCRM3 and CCRM3 predictions, and presents the conclusions and future plans.

## 2. Derivation of MSCRM3

We derive MSCRM3 and residual corrections to it by starting with the nuclear Schrodinger equation:

$$\hat{H}_o |\Phi\rangle \equiv \left( \sum_{n=1}^{A} \frac{\hat{p}_n^2}{2M} + \hat{V} \right) |\Phi\rangle = E |\Phi\rangle, \quad (2)$$

where the interaction $\hat{V}$ is rotationally invariant, and $A$ is the mass number. Now we perform a dynamic (i.e., nucleon-generated) collective rotation of the wavefunction $|\Phi\rangle$ through the angle $\vec{\theta}$ to obtain the wavefunction $|\Psi\rangle$:

$$|\Phi\rangle = e^{-i\vec{\theta}\cdot\hat{\vec{J}}/\hbar} \cdot |\Psi\rangle \equiv e^{-Z} \cdot |\Psi\rangle, \quad (3)$$

where the function $|\Psi\rangle$ is determined below, and the three rotation angles $\vec{\theta}$ and hence the three components of the angular-momentum operator $\hat{\vec{J}}$ describe rotations about the three space-fixed frame axes[1]. Since $\vec{\theta}$ is chosen to be a vector function of the particle coordinates, the rotation operator $e^{-Z}$ in Eq. (3) is rotationally invariant, and hence the wavefunctions $|\Phi\rangle$ and $|\Psi\rangle$ are eigenfunctions of the angular momentum and are isotropic. Inserting $|\Phi\rangle$ in Eq. (3) into Eq. (2), we obtain the Schrodinger equation:

---

[1] We use the space-fixed-frame-axis components of the rotation-angle $\vec{\theta}$ and angular-momentum operator vectors because they simplify the analysis and yields a transformed Hamiltonian that is a function symmetric in the square of the angular momentum components along the space-fixed-frame axes, whereas the Euler angles about the body-fixed-frame axes yield a Hamiltonian that is complicated asymmetric function of the angular momentum components. However, since $Z$ in Eq. (3) is rotationally invariant, we may also use the components of $\vec{\theta}$ and $\hat{\vec{J}}$ along body-fixed axes as is done below.

$$\hat{H}|\Psi\rangle \equiv e^Z \cdot \hat{H}_o \cdot e^{-Z} \cdot |\Psi\rangle$$
$$= \left(\frac{1}{2M}\sum_{n=1}^{A} e^Z \cdot \hat{p}_n^2 \cdot e^{-Z} + \hat{V}\right)|\Psi\rangle = E|\Psi\rangle . \quad (4)$$

We note that, in Eq. (4), $\vec{\theta}$ and $\hat{p}_n^2$ do not commute. The transformed Hamiltonian $\hat{H}$ in Eq. (4) is observed to be rotationally invariant since:

$$\left[\hat{\vec{J}}, e^Z \cdot \hat{p}_n^2 \cdot e^{-Z}\right] = e^Z\left[\hat{\vec{J}}, \hat{p}_n^2\right]e^{-Z} = 0$$
$$\left[\hat{\vec{J}}, \hat{V}\right] = 0 \quad (5)$$

The invariance in Eq. (5) is one of the major advantages of the rotation operator $e^Z$ over other forms of the rotation operators used in the literature.

We evaluate $e^Z \cdot \hat{p}_n^2 \cdot e^{-Z}$ in Eq. (4) using the following expansion in powers of $Z$:

$$e^Z \cdot \hat{p}_n^2 \cdot e^{-Z} = \hat{p}_n^2 + \left[Z, \hat{p}_n^2\right] + \frac{1}{2!}\left[Z, \left[Z, \hat{p}_n^2\right]\right] + \ldots \quad (6)$$

Using the rotational invariance of $Z$ and $\hat{p}_n^2$ in Eq. (5), we can easily evaluate the terms in Eq. (6) and show that the terms in powers of $Z$ higher than two vanish since $\vec{\theta}$ is a function of only the nucleon co-ordinates. Inserting these results into Eq. (4), we obtain:

$$\hat{H} = \hat{H}_o + \frac{i\hbar}{2M} \cdot \sum_{n=1}^{A}\sum_{k=1}^{3} \nabla_n^2 \theta_k \cdot \hat{J}_k$$
$$- \frac{1}{M} \cdot \sum_{n=1}^{A}\sum_{k=1}^{3} \left(\vec{\nabla}_n \theta_k \vec{\hat{p}}_n\right) \cdot \hat{J}_k \quad (7)$$
$$+ \frac{1}{2M} \cdot \sum_{n=1}^{A}\sum_{k=1}^{3} \vec{\nabla}_n \theta_k \cdot \vec{\nabla}_n \theta_l \cdot \hat{J}_k \cdot \hat{J}_l$$

Since each of the terms in Eq. (7) is rotationally invariant, we transform the components of each vector quantities in Eq. (7) to the intrinsic (i.e., body-fixed) axes. To express the third term in Eq. (7), which represents the coupling between the rotation and intrinsic motions, in terms of $\hat{\vec{J}}$, we use the following rigid-flow velocity-field prescription for the rotation angles $\vec{\theta}$:

$$\frac{\partial \theta_l}{\partial x_{nj}} = - \sum_{k \neq l \neq j=1}^{3} {}_l\chi_{jk} x_{nk}, \quad {}_l\chi_{jk} = -{}_l\chi_{kj} \quad (8)$$
$$(l, j, k = 1, 2, 3)$$

where each of the three $3\times 3$ matrices ${}_l\chi$ is real and anti-symmetric. The prescription in Eq. (8) renders $\hat{H}$ in Eq. (4) dependent on only $\hat{J}_k^2$ [2]. The non-zero elements of the matrices ${}_l\chi$ are determined by choosing $\theta_k$ and $\hat{J}_k$ to be a canonically conjugate pair, i.e., $\left[\theta_k, \hat{J}_k\right] = i\hbar$, which together with Eq. (8) yields:

$${}_l\chi_{jk} = \frac{1}{\hat{\mathscr{I}}_l}, \quad \hat{\mathscr{I}}_l \equiv \sum_{n=1}^{A}\left(x_{nj}^2 + x_{nk}^2\right),$$
$$j, k, l = 1, 2, 3 \text{ are in cyclic order} \quad (9)$$

where $M\hat{\mathscr{I}}_l$ is the $l^{th}$ principal-axis component of the rigid-flow moment of inertia tensor in the body-fixed frame[3]. We observe that, for the choice in Eqs. (8) and (9), we obtain the result:

$$\left[\theta_k, \hat{J}_l\right] = i\hbar \frac{Q_{kl}}{\hat{\mathscr{I}}_k} \approx 0 \, (k \neq l), \quad (10)$$

since $Q_{lk} \equiv \sum_{n=1}^{A} x_{nl} \cdot x_{nk}$ (which is the $kl^{th}$ component of the quadrupole-moment tensor) for $k \neq l$ is very small compared to $\hat{\mathscr{I}}_k$. (However, there is no need to know the action of $\hat{J}_l$ on $\theta_k$ for $k \neq l$ in the derivation of any of the equations in this article.) Inserting Eqs. (8) and (9) into Eq. (7), we obtain:

---

[2] We note that the choice of the rotation angles (for example, rigid or irrotational, etc. flow) is arbitrary and does not change the energy. It changes the term representing the coupling of intrinsic and rotational motions, and hence the division of the Hamiltonian between rotation and intrinsic motions. For an angle prescription different from the rigid-flow, higher order terms in $\hat{J}_k$ coupled to other intrinsic (such as shear) operators would appear in $\hat{H}$ in Eq. (4). Therefore, the collective rigid-flow velocity-field prescription in Eq. (8) accounts in an optimum way for the interaction between the angular momentum (i.e., collective rotation) and other types of operators (intrinsic motion), which is manifested in the appearance of the rigid-flow kinematic moment of inertia $\hat{\mathscr{I}}_k$ in Eqs. (9) and (11).

[3] The rigid-flow prescription for $\theta_k$ in Eqs. (8) and (9) is a collective analogue of the Birbrair's single-particle $\theta_n$ [26]. $\theta_n$ has continuous second-order mixed derivatives in any spatial region that excludes the rotation $x$ axis along which $\theta_n$ is singular. Whereas $\theta_k$ has discontinuous second-order mixed derivatives. The difference in the discontinuity between $\vec{\nabla}_n \theta_k$ and $\vec{\nabla}_n \theta_n$ arises because of the many-body nature of $\left(\hat{\mathscr{I}}_k\right)^{-1}$ in Eq. (9). However, the discontinuity in $\vec{\nabla}_n \theta_k$ is small because the expectation of $\hat{\mathscr{I}}_k$ is a large number. On the other hand, for a rigid-body motion, with a constant angular velocity, $\vec{\nabla}_n \theta_{rig}$ is inherently discontinuous. Nevertheless, $\theta_{rig}$ is commonly used in the analysis of a rigid-body motion [26,27] with impunity because the second-order mixed derivatives of $\theta_{rig}$ do not appear in any of the analysis equations. Similarly, the discontinuity in $\vec{\nabla}_n \theta_k$ is of no consequence in the derivation of MSCRM3 because the second-order mixed derivatives of $\theta_k$ do not appear anywhere in the derivation of the equations.

$$\hat{H} \equiv \hat{H}_o - \sum_{k=1}^{3} \frac{1}{2M\hat{\mathcal{I}}_k} \cdot \hat{J}_k^2 - \sum_{k \neq l=1}^{3} \frac{\hat{Q}_{lk}}{2M\hat{\mathcal{I}}_l \hat{\mathcal{I}}_k} \cdot \hat{J}_l \cdot \hat{J}_k, \quad (11)$$

where $\hat{H}$ is rotationally and time-reversal invariant.

Inserting Eq. (11) into Eq. (4), we obtain the following Schrodinger equation for a triaxial rotation of a microscopic quantum triaxial rotor:

$$\hat{H}|\Psi\rangle \equiv \left( \hat{H}_o - \sum_{k=1}^{3} \frac{\hat{J}_k^2}{2M\hat{\mathcal{I}}_k} - \sum_{k \neq l=1}^{3} \frac{\hat{Q}_{lk}}{2M\hat{\mathcal{I}}_l \hat{\mathcal{I}}_k} \cdot \hat{J}_l \cdot \hat{J}_k \right) |\Psi\rangle = E|\Psi\rangle. \quad (12)$$

Eq. (12) shows that $|\Psi\rangle$ is not an eigenfunction of $\hat{H}_o$ but rather it is an eigenstate of the reduced Hamiltonian $\hat{H}$.

One can easily show that:

$$\left[\theta_k, \hat{H}\right] = \frac{-i\hbar}{\hat{\mathcal{I}}_k} \left( \frac{\hat{Q}_{kl}}{\hat{\mathcal{I}}_l} \hat{J}_l + \frac{\hat{Q}_{kk'}}{\hat{\mathcal{I}}_{k'}} \hat{J}_{k'} \right) \approx 0, \quad (13)$$
$$(k \neq l \neq k' = 1,2,3)$$

since $\hat{Q}_{lk}$ for $k \neq l$ is very small compared to $\hat{\mathcal{I}}_l$. The result in Eq. (13) indicates that $\hat{H}$ in Eq. (12) is nearly independent of $\hat{J}_l$ and hence $\hat{H}$ is optimally or ideally intrinsic. The Hamiltonian $\hat{H}$ is also time-reversal invariant because it is quadratic in $\hat{J}_k$, and is rotationally invariant since the rotation operator $e^Z$ in Eq. (3) is rotationally invariant.

The inverse of the rigid-flow kinematic moment of inertia $\hat{\mathcal{I}}_l$ is a many-body operator. To render Eq. (12) solvable, we replace $\hat{\mathcal{I}}_l$ in Eq. (12) by its expectation value:

$$\mathcal{I}_l \equiv \langle \Psi | \hat{\mathcal{I}}_l | \Psi \rangle. \quad (14)$$

This is a reasonably good approximation since $\hat{\mathcal{I}}_l$ is a relatively large number and varies little and gradually with the angular momentum quantum number $J$ [4]. Eq. (12) then becomes:

$$\hat{H}|\Psi\rangle \equiv \left( \hat{H}_o - \sum_{k=1}^{3} \frac{\hat{J}_k^2}{2M\mathcal{I}_k} - \sum_{k \neq l=1}^{3} \frac{\hat{Q}_{lk}}{2M\mathcal{I}_l \mathcal{I}_k} \cdot \hat{J}_l \cdot \hat{J}_k \right) |\Psi\rangle = E|\Psi\rangle. \quad (15)$$

We now apply HF variational method to Eq. (15) and obtain:

$$\hat{H}|\Psi\rangle = \left[ \hat{H}_{crHF} + \hat{V}_{res} - \sum_{k=1}^{3} \frac{(\hat{J}_k^2)_{res}}{2M\mathcal{I}_k} \right] |\Psi\rangle = E|\Psi\rangle, \quad (16)$$

where $\hat{H}_{crHF}$ is the HF mean-field independent-particle part of $\hat{H}$ and satisfies the Schrodinger equation:

$$\hat{H}_{crHF}|\Psi_{crHF}\rangle$$
$$\equiv \left( \hat{H}_{oHF} - \vec{\Omega} \cdot \vec{J} - \sum_{l \neq k=1}^{3} \frac{Q_{lk}}{\mathcal{I}_k} \cdot \Omega_l \cdot \hat{J}_k \right.$$
$$\left. - \frac{1}{2} \sum_{l \neq k=1}^{2} \Omega_k \cdot \Omega_l M \hat{Q}_{lk} \right) |\Psi_{crHF}\rangle \quad (17)$$
$$\equiv E_{crHF}|\Psi_{crHF}\rangle$$

where $Q_{lk} \equiv \langle \Psi_{crHF} | \hat{Q}_{lk} | \Psi_{crHF} \rangle \equiv \langle \hat{Q}_{lk} \rangle$. $\hat{H}_{oHF}$ in Eq. (17) is the HF mean-field independent-particle part of $\hat{H}_o$. $\vec{\Omega} \cdot \vec{J}$ in Eq. (17) is the single-particle direct HF mean-field part of $\sum_{k=1}^{3} \frac{\hat{J}_k^2}{2M\mathcal{I}_k}$, and $\vec{\Omega}$ is the angular-velocity vector with the magnitude $\Omega$ and

---

[4] In fact, we note that a variation in $|\Psi\rangle$ in $1/\mathcal{I}_l$ results in the factor $1/\mathcal{I}_l^2$, which is very small compared to $1/\mathcal{I}_l$. The approximation is even more appropriate for heavier nuclei (with higher mass numbers) since the kinematic moment of inertia $\mathcal{I}_l$ is even larger and a change in the configuration of a few nucleons will not change $\mathcal{I}_l$ by much. The approximation is also valid in nuclei exhibiting CAP (Coriolis Anti-Pairing) and RAL (Re-Alignment) of the spins of a paired nucleon (quasi-particle) and the resulting back bending in the dynamic (and not in the kinematic) moment of inertia (plotted versus the square of the angular velocity $\vec{\Omega}$) and band-crossing [28]. The CAP-RAL-band-crossing phenomenon changes significantly the dynamic moment of inertia (defined as by $\mathcal{I}_d \equiv 2\hbar^2 (J-1)/\Delta E_J$, where $\Delta E_J$ is the excitation energy) but not the kinematic moment of inertia because the latter does not involve the nucleon dynamics (interaction) whereas the former does. Eq. (18) shows that the large change in the angular velocity arises from a corresponding change in the angular momentum and not in $\mathcal{I}_l$. However, the clear connection between the cranked wavefunction $|\Psi_{crHF}\rangle$ in the cranked Schrodinger Eq. (17) and the corresponding angular-momentum eigenstate $|\Psi\rangle$ in Eq. (16) would allow us to relate the back-bending condition to a definite angular momentum quantum number. $^{48}$Cr serves a good case to test this connection.

The predictions of MSCRM3 presented in this article have demonstrated that the microscopic angular velocity can have significant impact on the stability of a given cranked rotational state and its transition to other rotational states, and possibly altering the excitation energy and angular momentum at band termination. It is then conceivable that MSCRM3 may also affect the angular velocity and momentum in the back-bending/band-crossing region studied in [29,30]. These nuclear rotation features will be studied in future applications of MSCRM3.

orientation $(\theta, \phi)$, and is defined by its three Cartesian components:

$$\vec{\Omega} \equiv (\Omega_1, \Omega_2, \Omega_3)$$
$$\equiv \Omega(\sin\theta\cos\phi, \sin\theta\sin\phi, \cos\theta). \quad (18)$$
$$= \left(\frac{\langle\hat{J}_1\rangle}{M\mathscr{I}_1}, \frac{\langle\hat{J}_2\rangle}{M\mathscr{I}_2}, \frac{\langle\hat{J}_3\rangle}{M\mathscr{I}_3}\right)$$

$(\hat{J}_k^2)_{res}$ in Eq. (16) is the residual of the square of the angular-momentum operator given by:

$$(\hat{J}_k^2)_{res} \equiv \hat{J}_k^2 - 2\langle\Psi_{crHF}|\hat{J}_k|\Psi_{crHF}\rangle \cdot \hat{J}_k. \quad (19)$$

For the residual of nuclear interaction $\hat{V}_{res}$ in Eq. (16), we may use the commonly-used separable effective quadrupole-quadrupole (long range) interaction.

In Eq. (16) we do not show the residual of the three-body operator $-\sum_{l \neq k=1}^{2} \frac{\hat{Q}_{lk}}{2M\mathscr{I}_l\mathscr{I}_k} \cdot \hat{J}_l \cdot \hat{J}_k$ in Eq. (15) because it is negligibly small, complicated, and is not used in this article[5].

Eq. (17) shows that the cranking model wavefunction $|\Psi_{crHF}\rangle$ is an eigenstate of the cranking Hamiltonian $\hat{H}_{crHF}$, which is the HF mean-field part of the intrinsic rotor Hamiltonian $\hat{H}$ in Eq. (15). Since for open-shell nuclei, $|\Psi_{crHF}\rangle$ describes an anisotropic nucleon spatial distribution, the residual terms in Eq. (16) transforms $|\Psi_{crHF}\rangle$ into $|\Psi\rangle$, which is isotropic and an eigenstate of the angular momentum. Generally, we may expect these residuals to be relatively small to render the HF wavefunction $|\Psi_{crHF}\rangle$ useful and meaningful. Therefore, we approximate $|\Psi\rangle$ by $|\Psi_{crHF}\rangle$ and hence CCRM3 ignores the residual terms in Eq. (16) and the normally small quadrupole terms in Eq. (17) (which may become significant for wobbly and triaxial rotations at high spins and in back-bending regions, where $\Omega_k$ and $\hat{J}_k$ undergo large changes). When these terms are neglected, Eq. (17) becomes:

$$\hat{H}_{crHF}|\Psi_{crHF}\rangle = (\hat{H}_{oHF} - \vec{\Omega}\cdot\vec{\hat{J}})|\Psi_{crHF}\rangle$$
$$= E_{crHF}|\Psi_{crHF}\rangle \quad . \quad (20)$$

Since $\vec{\Omega}$ and $\vec{\hat{J}}$ change sign on time reversal, $\hat{H}_{crHF}$ is time-reversal, $D_2$, and signature invariant. The $D_2$ and signature invariances render the MSCRM3 wavefunction a linear superposition of either even or odd angular-momentum eigenstates, in contrast to a CCRM3 wavefunction that is a superposition of even and odd angular-momentum eigenstates. This result resolves a conceptual problem in comparing predicted and measured nuclear spectra identified in [17]. Time-reversal invariance renders a MSCRM3 state and its time-reversed companion degenerate possibly impacting the consequences of nucleon pairing interaction.

MSCRM3 Eq. (20) is identical in form to CCRM3 Eq. (1) except for the microscopic angular-velocity $\vec{\Omega}$. As is done for CCRM3 Eq. (1), we now present the remaining equations needed to solve iteratively Eq. (20) using a self-consistent deformed harmonic oscillator potential $M\sum_{k=1}^{3}\omega_k^2 x_k^2/2$ [6]. Taking the expectation of $\hat{H}_{crHF}$ in Eq. (20), we obtain the intrinsic energy:

$$E_{int} \equiv E_{crHF} = \langle\hat{H}_{oHF} - \vec{\Omega}\cdot\vec{\hat{J}}\rangle = \hbar\sum_{k=1}^{3}\alpha_k\Sigma_k$$
$$\Sigma_k \equiv \sum_{n_k=1}^{n_{kf}}(n_k+1) \quad , \quad (21)$$

and the rotational-band excited-state ($E_J$) and excitation ($\Delta E_J$) energies are then determined from:

$$E_J \equiv \langle\hat{H}_o\rangle \equiv \langle\hat{H}_{crHF} + \vec{\Omega}\cdot\vec{\hat{J}}\rangle = E_{int} + \vec{\Omega}\cdot\langle\vec{\hat{J}}\rangle \quad , \quad (22)$$
$$\Delta E_J \equiv E_J - E_{J=0}$$

where $\alpha_k$ are the three normal-mode frequencies, which are determined from the equations-of-motion for $\hat{H}_{crHF}$ in Eq. (20) to be the three roots of the characteristic equation [15]:

$$\alpha^6 - A_o\alpha^4 + B_o\alpha^2 - C_o = 0. \quad (23)$$

where $A_o$, $B_o$, and $C_o$ are functions of $\omega_k^2$ and $\Omega_k^2$.

---

[5] This term, which is normally small, but may become important for wobbly and triaxial rotations at high spins and in back-bending regions where $\Omega_k$ and $\hat{J}_k$ undergo large changes.

[6] Nucleon pairing and spin-orbit interactions are not included in the preliminary and exploratory MSCRM3 application and calculations performed in this article to provide physical insight on phenomena and kinds of changes in the predicted rotational properties generated by microscopic angular velocity. These phenomenon and property changes may provide useful information and guidance in a planned detailed calculation using MSCRM3 with realistic nuclear interactions. The results predicted in this article are also compared with those of previous CCRM3 calculations [2-5,7,8,10-19], which also did not use pairing and spin-orbit interactions. However, in the $^{20}$Ne case, where pairing is insignificant, we use spin-orbit interaction and residuals of the square of the angular momentum and two-body interactions to match the predicted and measured excitation energy after quenching of wobbly rotation.

The expectations of $\hat{J}_k$ and $\sum_{n=1}^{A} x_{nk}^2$ are then given by (using Feynman's theorem [31]):

$$\left\langle \hat{J}_k \right\rangle = -\frac{\partial E_{\text{int}}}{\partial \Omega_k'} = -\hbar \sum_{l=1}^{3} \frac{\partial \alpha_l}{\partial \Omega_k'} \Sigma_l$$

$$\frac{M}{2}\left\langle \sum_{n=1}^{A} x_{nk}^2 \right\rangle \equiv \frac{M}{2}\left\langle x_k^2 \right\rangle = \frac{\partial E_{\text{int}}}{\partial \omega_k^2} = \hbar \sum_{l=1}^{3} \frac{\partial \alpha_l}{\partial \omega_k^2} \Sigma_l \quad . \quad (24)$$

Eqs. (24) must also satisfy the constant nuclear-volume ($\omega_1 \omega_2 \omega_3 = \omega_o^3$) and the density-potential shapes self-consistency conditions. This implies, in the HF mean-field sense, a minimization of $E_{\text{int}}$ with respect to the oscillator frequencies $\omega_k^2$, yielding the self-consistency conditions:

$$\omega_1^2 \left\langle x^2 \right\rangle = \omega_2^2 \left\langle y^2 \right\rangle = \omega_3^2 \left\langle z^2 \right\rangle , \quad (25)$$

where $\omega_o$ is the isotropic nuclear harmonic-oscillator frequency given by $\hbar \omega_o \equiv P_o A^{-1/3}$.

In Eq. (20), the wavefunction $|\ \rangle \equiv |\Psi_{\text{crHF}}\rangle$ is required to yield the angular momentum constraint (as in CCRM3):

$$\left\langle \hat{J}^2 \right\rangle \equiv \left\langle \hat{J}_1 \right\rangle^2 + \left\langle \hat{J}_2 \right\rangle^2 + \left\langle \hat{J}_3 \right\rangle^2 = \hbar^2 J(J+1)$$
$$\text{for a triaxial rotation}$$
$$, \quad (26)$$
$$\left\langle \hat{J}^2 \right\rangle \equiv \left\langle \hat{J}_1 \right\rangle^2 + \left\langle \hat{J}_2 \right\rangle^2 + \left\langle \hat{J}_3 \right\rangle^2 = \hbar^2 J^2$$
$$\text{for a principal-axis rotation}$$

where $J$ is the desired (or externally imposed) angular-momentum quantum number. However, unlike that in CCRM3, the $J_{cal}$ value calculated from $\hbar J_{cal} = \left\langle \hat{J}^2 \right\rangle^{1/2}$ for principal-axis rotation or from $\hbar \sqrt{J_{cal}(J_{cal}+1)} \equiv \left\langle \hat{J}^2 \right\rangle^{1/2}$ for triaxial rotation is generally different from $J$ because of the interaction between $\left\langle \hat{J}_k \right\rangle$ and $\Omega_k$ generated by Eq. (18) (or equivalently by (28) given below) and the resulting rotational relaxation of the intrinsic system, causing $J_{cal}$ determined from Eq. (26) to be either smaller or larger than the desired or imposed $J$ value.

The orientation angles $(\theta, \phi)$ and the magnitude $\Omega$ of $\vec{\Omega}$ are readily determined from Eq. (18), and are given by:

$$\tan \phi = \frac{\left\langle \hat{J}_2 \right\rangle \mathcal{I}_1}{\left\langle \hat{J}_1 \right\rangle \mathcal{I}_2}, \quad \tan \theta = \frac{\left\langle \hat{J}_1 \right\rangle \mathcal{I}_3}{\left\langle \hat{J}_3 \right\rangle \mathcal{I}_1} \sqrt{1 + \tan^2 \phi} , \quad (27)$$

$$\Omega^2 = \Omega_1^2 + \Omega_2^2 + \Omega_3^2$$
$$= \frac{\left\langle \hat{J}_1 \right\rangle^2}{M^2 \mathcal{I}_1^2} + \frac{\left\langle \hat{J}_2 \right\rangle^2}{M^2 \mathcal{I}_2^2} + \frac{\left\langle \hat{J}_3 \right\rangle^2}{M^2 \mathcal{I}_3^2} , \quad (28)$$
$$= (T_1 + T_2 + T_3)/3$$

where $\mathcal{I}_k$ is given in Eqs. (9) and (14), and $\left\langle \hat{J}^2 \right\rangle$ is defined in Eq. (26), and:

$$T_j \equiv \frac{\left\langle \hat{J}^2 \right\rangle}{M^2 \mathcal{I}_j^2} + \left(\frac{1}{M^2 \mathcal{I}_k^2} - \frac{1}{M^2 \mathcal{I}_j^2}\right)\left\langle \hat{J}_k \right\rangle^2$$
$$+ \left(\frac{1}{M^2 \mathcal{I}_l^2} - \frac{1}{M^2 \mathcal{I}_j^2}\right)\left\langle \hat{J}_l \right\rangle^2 .$$

($j, k, l = 1, 2, 3$ are in cyclic order)

## 3. MSCRM3 and CCRM3 predictions

MSCRM3 Eqs. (14), (20), (21), and (23)-(28) are solved iteratively for each desired $J$ value and for ground-state nucleon configurations of the nuclei $^{20}$Ne, $^{24}$Mg, and $^{28}$Si. (An identical set of CCRM3 equations are similarly solved except that in Eqs. (27) $\mathcal{I}_k$ is unity and Eq. (28) is not used, but instead $\Omega$ is adjusted to give the desired $J$ value.) We use a linear-mixing method [32], where the input and output values of the variables in each iteration step are linearly mixed and used as input in the next iteration step, to obtain steady converged solutions. From Eq. (22), the values of $E_J$ and $\Delta E_J$ are computed for each desired value of $J$ and the initial (boundary) values $\phi_o$ and $\theta_o$ of $\phi$ and $\theta$. The initial values $\phi_o$ and $\theta_o$ determine the nature of the subsequent rotation [7].

---

[7] For a given value of the imposed angular momentum quantum number $J$ in the conventional or microscopic cranking model for uniaxial rotation (i.e., CCRM1 or MSCRM1), the angular-velocity-vector angles $\phi$ and $\theta$ take on only the constant values $\phi = 0$ and $\theta = \pi/2$. Therefore, there is just one type of rotation, namely a uniform rotation about the principal axis $x$, which is assumed to yield the minimum excitation (or yrast) energy. For triaxial rotation in either CCRM3 or MSCRM3 at each $J$, there can be many rotation types: uniform rotation about either of the three principal axis, planar (either uniform or tilted) rotation, and triaxial rotation where none of the angular-momentum components is zero in any chosen quadrant among the eight coordinate system quadrants. For a given $J$, each of these numerous types of rotation is conveniently labelled (or distinguished) by a given set of initial values $\phi_o$ and $\theta_o$ of $\phi$ and $\theta$, and by the energy $E_J$ or $\Delta E_J$, which have also been used in previous analyses [17,18,33]. For a given $J$, two or more of these rotation types (i.e., different values of $\phi_o$ and $\theta_o$) may have the same (degenerate) or different excitation energies $\Delta E_J$, depending on the nature of the nucleon interaction used. The type or types of rotation that yields or yield the minimum excitation energy is or are of particular physical interest and may labelled by ($J$, $\phi_{o\min}$, $\theta_{o\min}$

Since the CCRM3 and MSCRM3 governing equations are similar in form, their predictions are similar. But there are also cases that show significant differences because of the microscopic angular velocity in MSCRM3. We now present the latter cases and also compare these preliminary and exploratory results with those of the CCRM3 predictions reported in ~~this section and in~~ [7,8,10,12-19], noting that these preliminary and exploratory results provide useful information and guidance on the more complicated types of rotational phenomena generated when pairing and spin-orbit interactions are included in the models.

CCRM3 always predicts uniform rotation, i.e., the $\vec{\Omega}$ and $\langle \vec{\hat{J}} \rangle$ are parallel vectors. In contrast, MSCRM3 predicts wobbly (tilted-axis) rotation (compare Eq (27) for the two models).

The CCRM3 and MSCRM3 rotational states are inherently unstable because the self-consistency condition in Eq. (25) causes $E_J$ in Eq. (22) to increase and $E_{int}$ in Eq. (21) to decrease with $J$. This generates a destabilizing positive-feedback between $\Omega_k$ and $\langle \hat{J}_k \rangle$ (consistently with the first of Eq. (24)) so that a decrease in $\langle \hat{J}_k \rangle$ requires $\Omega_k$ to decrease, which in turn causes $\langle \hat{J}_k \rangle$ to decrease further until $\langle \hat{J}_k \rangle$ is reduced to zero, unless this free fall is prevented by some constraint such as that in Eq. (26)[8].

For the constraint in Eq. (26), a MSCRM3 or CCRM3 rotational state does not totally collapse rotationally (i.e., all three components $\langle \hat{J}_k \rangle$ simultaneously decrease to zero). For this reason and as seen in Fig. 1, $\langle \hat{J}_3 \rangle$ normally decreases to zero, resulting in a planar uniform rotation in the $x$-$y$ plane in CCRM3 and a planar wobbly rotation in the $x$-$y$ plane in MSCRM3. Then, any small decrease in, for example, $\langle \hat{J}_1 \rangle$ results in a monotonic decrease in $\langle \hat{J}_1 \rangle$ and an increase in $\langle \hat{J}_2 \rangle$, resulting in a uniform rotation about the principal $y$ axis, and similarly for a small decrease in $\langle \hat{J}_2 \rangle$. An exception is when the planar rotation is about the $\phi_o = 45°$ line in the $x$-$y$ plane in which case the planar uniform rotation is stable in CCRM3 but not the planar wobbly rotation in MSCRM3, as in the $^{20}$Ne case discussed below.

From the above results, we conclude that CCRM3 predicts only uniform rotation (about the 45° line in the $x$-$y$ plane or about the $x$ or $y$ axis with $\theta = 90°$), and it does not predict any triaxial rotations. This may or may not be the case in MSCRM3.

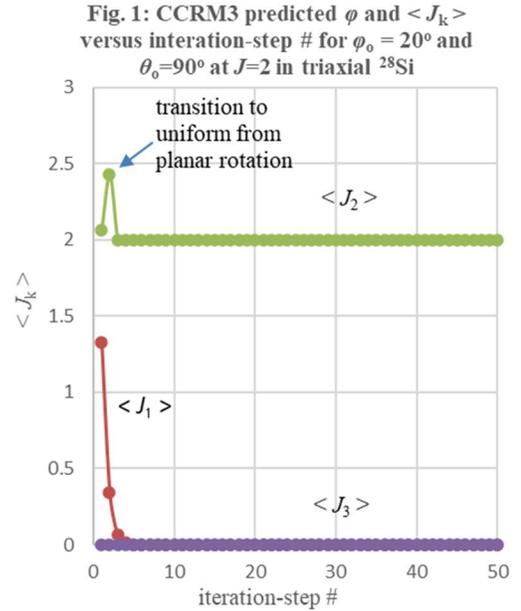

Fig. 1: CCRM3 predicted $\varphi$ and $<J_k>$ versus interation-step # for $\varphi_o = 20°$ and $\theta_o = 90°$ at $J=2$ in triaxial $^{28}$Si

The self-consistency between $\Omega_k$ and $\langle \hat{J}_k \rangle$ in MSCRM3 also allows the intrinsic system to rotationally relax after the imposed $J$ in the first iteration step is removed and replaced by Eq. (28). This rotational relaxation yields a value of $J_{cal}$ (defined above) that is smaller or larger than $J$. In contrast, CCRM3 forces the intrinsic system to accept the imposed value $J$ that may be inconsistent with the nucleonic motion.

---

, and $\Delta E_{J\min}$ ). The values of $\phi_o$ and $\theta_o$ can be evaluated from Eq. (27) given the expectation values of the angular-momentum and quadrupole-moment components in Eq. (24) and $\mathcal{J}$ in Eq. (9).

Therefore, $\phi_o$ and $\theta_o$ become (indirectly) observables and hence can be measured. In fact, [33] states "Note that azimuthal ($\phi$) and polar ($\theta$) angles are measured, respectively, from the 1-3 plane and the 1-axis.". The $^{20}$Ne case studied in Fig. 2 is a good example of the above narrative. For $\phi_o = 0$, $\theta_o = \pi/2$ and $J=8$, both CCRM3 and MSCRM3 predict nearly the same excitation energy $\Delta E_{J=8}$.

But $\Delta E_{J=8}$ does not correspond to the minimum measured energy. The minimum energy $\Delta E_{J=8\min}$ is the excitation energy predicted by MSCRM3 in Fig. 2 at $\phi_o = 45°$, $\theta_o = \pi/2$ and is a consequence of the quenching of the wobbly rotation at $J < 8$ and the resulting transition to uniform principal $x$-axis rotation at $J=8$. The quenching of wobbly rotation (i.e., the transition to uniform rotation) is the cause of the measured reduced energy-level spacing between $J=6$ and 8.

[8] This result implies that the rotational instability occurs for any self-consistent mean-field potential and not just for the oscillator potential.

For prolate $^{20}$Ne (axially-symmetric about $z$ axis), $\phi_o = 45°$, and $\theta_o > 0°$, MSCRM3 predicts a planar wobbly rotation in the $x$-$y$ plane at $J_{cal} \lesssim 6$. At $J_{cal} = 7.76$, $^{20}$Ne becomes triaxial (due to the interaction between centrifugal stretching and incompressibility condition), and the fluctuations in $\langle \hat{J}_2 \rangle$ are amplified reducing it to zero. This causes a transition from planar in the x-y plane to uniform rotation about the $x$ axis (i.e., quenching of the wobbly rotation) and the switching from the constraint $\langle \hat{J}_{cal}^2 \rangle = \hbar^2 J(J+1)$ to $\langle \hat{J}_{cal}^2 \rangle = \hbar^2 J^2$. At $J_{cal} = 8$, $^{20}$Ne becomes axially symmetric about the $x$ axis and the rotational band terminates. The quenching of the wobbly rotation reduces the energy-level spacing between $J_{cal}$=6 and 8 and improves the agreement with the measured excitation energy as seen in Fig. 2[9]. The remaining discrepancy between the measured and MSCRM3-predicted excitation energies in Fig. 2 is partly due to the neglect of the spin-orbit interaction, which we have estimated to increase the predicted energy by 10%. The still remaining discrepancy is due to the neglect of the residuals of $\hat{J}_k^2$ and $\hat{V}$ in Eq. (16), which is estimated, using MSCRM1 for uniaxial rotation, Tamm-Dancoff approximation, and cranked 1-particle 1-hole basis states, to be 1 to 2 $MeV$ increasing the predicted excitation energy close to the measured excitation energy. In contrast, in CCRM3 the uniform planar rotation of $^{20}$Ne about the $\phi_o$=45° line in the x-y plane persists at all $J$ values with increasing excitation energy and no rotational-band termination, as seen in Fig. 2.

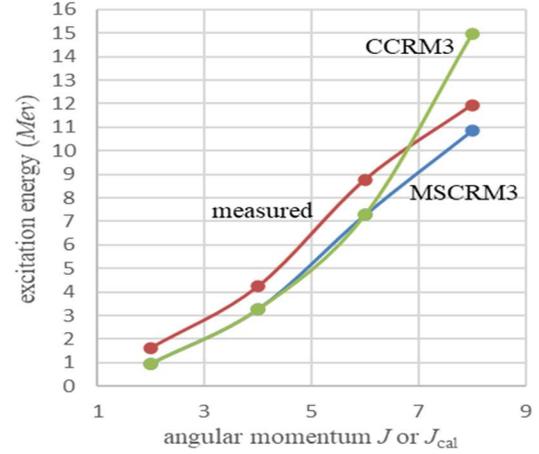

Fig. 2: Measured and MSCRM3 and CCRM3 predicted excitation energies versus $J$ or $J_{cal}$ for $\varphi_o = 45°$ in prolate $^{20}$Ne

Fig. 2 for $^{20}$Ne shows that, at $J$ or $J_{cal}$ =8, it is not the principal-axis (i.e., $\phi_o = 0°$, $\theta_o = \pi/2$) rotation but rather the planar wobbly (i.e., $\phi_o = 45°$, $\theta_o = \pi/2$) rotation that determines the minimum excitation energy (i.e., the yrast measured state and its energy). This result implies that, for a given nucleus and $J$ or $J_{cal}$, we need to explore all values of the pair $(\phi_o, \theta_o)$ to determine the yrast (i.e., minimum) excitation energy.

Fig. 3 shows that the quadrupole moments $Q$ ($Q \equiv Z \cdot e \cdot (2\langle z^2 \rangle - \langle x^2 \rangle - \langle y^2 \rangle)/A$), predicted by CCRM3 and MSCRM3 for $^{20}$Ne for $\phi_o = 45°$, $\theta_o = \pi/2$, decrease up to $J$=6 as does the measured quadrupole moment, (where the measured values are from [2,34,35]). Beyond $J$=6, the $Q$ predicted by MSCRM3 decreases (as does the measured $Q$) as $^{20}$Ne becomes progressively axially symmetric and the rotational band terminates at $J$=8. However, the $Q$ predicted by CCRM3 increases beyond $J$=6 because of the increasing centrifugal stretching at higher $J$ since the predicted rotational band does not terminate. It is noted in Fig. 3 that we obtain only a qualitative agreement between the measured and MSCRM3 predicted $Q$ because we have not included $\vec{l} \cdot \vec{s}$, $l^2$, and the residuals of the square of the angular momentum and two-body interactions in the quadrupole-moment calculation (refer to footnote 6 for more detail). These interactions alter the order of the single-particle state energies and spins versus nuclear deformation (as the Nilsson's model demonstrates). In any case, there are also large uncertainties in the measured $Q$.

---

[9] This reduction in energy-level spacing has been known [2,4] but has not been attributed to any specific phenomenon. MSCRM3 attributes this phenomenon to the quenching of wobbling at $J_{cal}$=8.

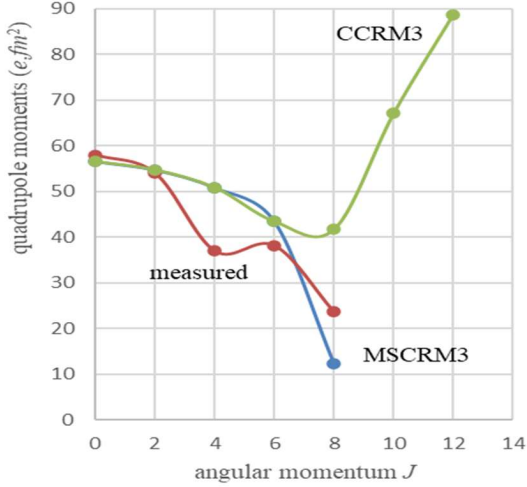

Fig. 3: MSCRM3 & CCRM3 predicted & measured quadrupole moments versus $J$ for $\varphi_o = 45°$ in $^{20}$Ne

For prolate $^{24}$Mg and $^{28}$Si, and $\phi_o=45°$, MSCRM3 predicts a wobbly planar rotation in the $x$-$y$ plane at respectively $J_{cal}<4$ and 8. At respectively $J_{cal}=4$ and 10, the planar rotation transitions to a uniform rotation about the $x$ or $y$ axis and the nuclei become triaxial. At respectively $J_{cal}=4$ and 12 the nuclei become axially symmetric about the $x$ or $y$ axis and the rotational band terminates. In contrast, CCRM3 predicts a planar uniform rotation about the 45° line in the $x$-$y$ plane of an axially symmetric (about $z$ axis) nucleus at all $J$ values and no band termination.

For triaxial $^{24}$Mg, $\phi_o=90°$, and $0°<\theta_o<90°$, MSCRM3 predicts $\langle \hat{J}_1 \rangle =0$ at all $J_{cal}$ values, and a steady planar rotation in the $y$-$z$ plane at $J_{cal}=2$ and 4, and a uniform steady rotation about the $y$ axis (i.e., $\langle \hat{J}_1 \rangle = \langle \hat{J}_3 \rangle =0$, and $\langle \hat{J}_2 \rangle \neq 0$) at $J=6$. At $J_{cal}=8$, the nucleus becomes axially symmetric about the $y$ axis and the rotational band terminates. In contrast, for $\phi_o=90°$ and $\theta_o=90°$, the nucleus rotates about the $y$ axis and remains triaxial at all $J_{cal}$ and the rotational band does not terminate. The above two cases show that a band termination and its $J_{cal}$ value can be sensitive to the $\theta_o$ value. In contrast, for $\phi_o=90°$, and $0°<\theta_o\leq 90°$, CCRM3 predicts a uniform rotation about the $y$ axis of a triaxial nucleus at all $J$ values, and the rotational band does not terminate.

For triaxial $^{24}$Mg, $\phi_o=89°$, and $\theta_o=54.7°$ at $J_{cal}=2$, MSCRM3 predicts that the rotational-state instability reduces $\langle \hat{J}_1 \rangle$ to near zero at the end of the iteration process resulting in a nearly planar triaxial rotation of a triaxial $^{24}$Mg in a plane close to the $y$-$z$ plane as seen in Figs. 4 and 5. Figs. (4) and (5) show respectively the values of the angular-momentum and quadrupole-moment components at each $J_{cal}$ value at the end of the iteration process. A similar triaxial rotation of triaxial $^{24}$Mg is predicted at $J_{cal}=4$. At $J=6$, the rotation becomes uniform along the $y$ axis and $^{24}$Mg remains triaxial. At $J_{cal}=8$, $^{24}$Mg remains triaxial and the uniform rotation transitions to a triaxial rotation, as seen in Figs. 4 and 5. At $J_{cal}=10$ the nuclear rotation and shape become almost axially symmetric about the $x$ axis, as seen in Figs. (4) and (5). At $J_{cal} J=12$, the nuclear rotation and shape become spherically symmetric and the rotational band terminates.

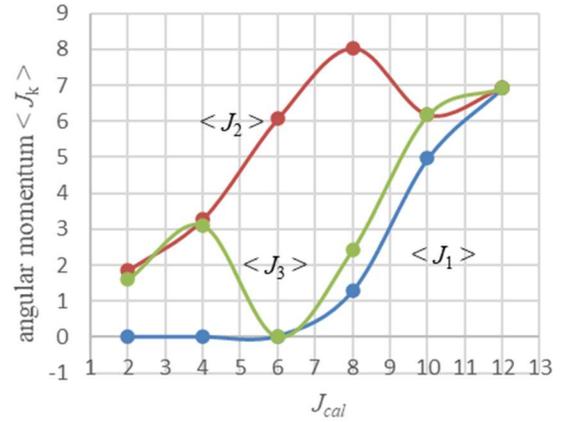

Fig. 4: MSCRM3 predicted $<J_k>$ versus $J_{cal}$ at end of iteration process for $\varphi_o=89°$ and $\theta_o=54.7°$ in triaxial $^{24}$Mg

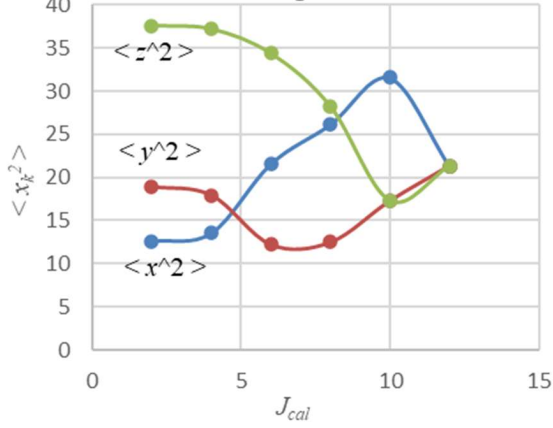

Fig. 5: MSCRM3 predicted quadrupole moments $<x_k^2>$ versus $J_{cal}$ for $\varphi_o=89°$ and $\theta_o=54.7°$ in triaxial $^{24}$Mg

For triaxial $^{24}$Mg, $\phi_o=89°$ and $\theta_o=90°$ at $J_{cal}=2$, MSCRM3 predicts, during the iteration process, a steady rotation about the $y$ axis with very small values of $\langle \hat{J}_1 \rangle$ and $\langle \hat{J}_3 \rangle$. These values begin to slowly increase to $J_{cal}=13$ and $10$ respectively because of competing effects of the three terms in the first of Eq. (24). At $J_{cal}=12$, $\langle \hat{J}_1 \rangle$ and $\langle \hat{J}_3 \rangle$ values become comparable to that of $\langle \hat{J}_2 \rangle$. Therefore, the uniform rotation about the $y$ axis transitions to a triaxial rotation of a triaxial nucleus and eventually to a spherically symmetric rotation of a spherical nucleus, at which point the rotational band terminates. This result contrasts with that for the case $\phi_o=89°$, and $\theta_o=54.7°$ described above. In contrast, CCRM3 predicts a uniform rotation of triaxial nucleus about the $x$ axis at $J=2,4,6$, and 8. At $J=8$, the nucleus becomes axially symmetric about the $x$ axis and the rotational band terminates.

For triaxial $^{24}$Mg, $21°\leq\phi_o<89°$, and $\theta_o=90°$, at $2\leq J_{cal}<12$, MSCRM3 predicts a steady uniform rotation about the $y$ axis of a triaxial nucleus. At $J_{cal}=12$, the nucleus and the rotation become spherically symmetric and the rotational band terminates, similarly to that for $\phi_o=89°$ and $\theta_o=90°$ at $J_{cal}=12$.

For triaxial $^{24}$Mg, $0°<\phi_o\leq20°$, and $\theta_o=90°$, MSCRM3 predicts no rotation of any kind at any $J_{cal}$; for $\phi_o=0°$ and $\theta_o=90°$, MSCRM3 predicts a uniform rotation about the $x$ axis at $2\leq J_{cal}<8$. At $J_{cal}=8$, the nucleus becomes axially symmetric about the $x$ axis and the rotational band terminates. In contrast, for $0°\leq \phi_o<90°$, and $0°<\theta_o\leq90°$, CCRM3 predicts a uniform rotation of triaxial nucleus about the $x$ axis at $J=2,4,6$, and 8. At $J=8$, the nucleus becomes axially symmetric about the $x$ axis and the rotational band terminates. For $\phi_o=90°$ and $0<\theta_o\leq90°$, CCRM3 predicts a uniform rotation about the $y$ axis of a triaxial nucleus at all $J$ values, and the rotational band does not terminate.

Shell-model calculation [36], experiments [37], and MSCRM3 prediction for $21°\leq\phi_o\leq89°$ and $0°<\theta_o\leq90°$ presented above show that the rotational band in $^{24}$Mg terminates at $J=12$ when the nucleus and rotation about the $y$ axis become spherically symmetric. In contrast, CCRM3 predicts that the rotational band terminates at $J=8$ when the nucleus becomes axially symmetric about the $x$ axis.

The above results also show that CCRM3 predicts only uniform principal-axis rotation of a triaxial $^{24}$Mg. This result may not be in conflict with the conclusions from the CCRM3 analysis in [19] because [19] used only one kind of nucleons.

For triaxial $^{28}$Si, neither MSCRM3 nor CCRM3 predict any physically meaningful collective rotation for any $J$, $J_{cal}$, $\phi_o$, and $\theta_o$ values.

For oblate $^{24}$Mg and $^{28}$Si, MSCRM3 and CCRM3 predict no steady stable collective rotation of any kind.

## 4. Summary, conclusions, and future plans

The conventional cranking model for uniaxial rotation is frequently used in nuclear structure studies. The model is semi-classical and phenomenological, because it uses a constant adjustable angular-velocity parameter. It is, therefore, time-reversal and $D_2$ non-invariant. To investigate the impact of a microscopic (dynamic) angular velocity on the results of such studies, a quantum microscopic self-consistent, cranking model Schrodinger equation for triaxial rotation (MSCRM3) is derived in two steps.

In the first step, we apply a dynamic rotationally-invariant exponential rotation operator to a rotationally-invariant nuclear Schrodinger equation to transform it into a rotationally and time-reversal invariant rotor Schrodinger equation. The rotor Hamiltonian is made optimally intrinsic by choosing the three rotation angles in the rotation operator to be given by a collective rigid-flow velocity-field prescription and to be canonically conjugate to the angular-momentum components.

In the second step, Hartree-Fock variation is applied to the rotor Schrodinger equation to obtain MSCRM3 Schrodinger equation with a Hamiltonian $\hat{H}_{crHF}$, which is the HF mean-field part of the rotor Hamiltonian, plus the residuals of the square of the angular-momentum operator and two-body interaction and other normally negligibly small correction terms. These residual terms may become important for wobbly and triaxial rotations at high spins and in back-bending regions where the angular velocity and momentum undergo large changes. Except for these correction terms, the MSCRM3 Schrodinger equation is identical in form to the conventional cranking-model equation for triaxial rotation (CCRM3) except that, in MSCRM3, the angular velocity emerges from the HF variation and is not a constant parameter but is a dynamical quantity given by the ratio of the expectations of the angular momentum and rigid-flow kinematic moment of inertia. As a consequence, $\hat{H}_{crHF}$ is time-reversal and $D_2$ invariant, rendering the MSCRM3 wavefunction a superposition of either even or odd angular-momentum eigenstates, in contrast to the CCRM3 wavefunction, which is a superposition of even and odd angular momentum eigenstates. This result resolves the difficulty of comparing the predicted and measured rotational spectra. Because $\hat{H}_{crHF}$ is time-reversal invariant, an eigenstate of $\hat{H}_{crHF}$ and its time-reversed state are degenerate, which may affect the consequences of the pairing interaction and the predicted rotational energy.

In this article, we present the results of a preliminary and exploratory calculation of rotational properties predicted by MSCRM3 and CCRM3 for the light nuclei $^{20}$Ne, $^{24}$Mg, and $^{28}$Si. For this purpose, we use the simple self-consistent deformed harmonic oscillator (mean-field) potential. This relatively simple calculation yields, in a transparent manner, physical insight and information on the types of rotational phenomena predicted by MSCRM3, and compare these with those predicted by CCRM3. These results serve as a guide in a planned calculation where realistic nuclear interactions will be used. For the simple oscillator potential, the MSCRM3 (and CCRM3) governing equations (or equivalently the three normal-mode frequencies of $\hat{H}_{crHF}$), coupled to those of the density-potential self-consistency and constant-volume conditions, are obtained in closed algebraic forms and solved using a simple iterative procedure.

Each predicted rotational mode is labelled by the values of angular-momentum quantum number $J$ imposed on the wavefunction, excitation energy, and by the two angular-velocity-vector polar angles at the start of the iteration process, as in previously reported CCRM3 analyses. Different polar-angle values may yield different excitation energies. Of particular interest are the polar-angle values that yield the minimum (yrast) rotational energy. The other angle values give excitations above the yrast trajectory.

A summary of the major differences between the predictions of MSCRM3 and CCRM3 for prolate $^{20}$Ne, $^{24}$Mg, and $^{28}$Si, and oblate and triaxial $^{24}$Mg, and $^{28}$Si is presented below.

CCRM3 predicts only uniform rotations whereas MSCRM3 predicts uniform, tilted-axis, wobbly, planar, and triaxial rotations.

CCRM3 and MSCRM3 predict that the coupling between the intrinsic and rotational motions generated by the density-potential self-consistency condition renders the rotational states unstable, causing them to decay to zero angular momentum unless prevented by an angular-momentum constraint on the wavefunctions.

In CCRM3, the rotational-state instability causes the initial triaxial rotation to transition to a planar uniform rotation in the $x$-$y$ plane and subsequently to a uniform rotation about either the $x$ or $y$ axis with a rotational-band termination at $J$=8,4, and 8 in prolate $^{20}$Ne, $^{24}$Mg, and $^{28}$Si respectively. An exception to this situation is when the initial rotation is axially symmetric, in which case it remains so at all $J$ values with no band termination. Therefore, CCRM3 does not predict any steady stable triaxial rotation in any of the nuclei $^{20}$Ne, $^{24}$Mg, and $^{28}$Si. These results may not be in conflict with the CCRM3 analysis result reported in [19] because [19] used only one kind of nucleons.

MSCRM3 predicts uniform principal-axis, planar non-uniform (i.e., wobbly or tilted-axis) rotations in the $x$-$y$ plane, and triaxial rotations and band termination at $J$=8 and 4, and 12 for prolate $^{20}$Ne, $^{24}$Mg, and $^{28}$Si respectively (and also at $J$=8 for initially non-uniform rotation in prolate $^{28}$Si). For planar wobbly initially axially-symmetric (i.e., along the 45$^\circ$ line in the $x$-$y$ plane) rotation in $^{20}$Ne, MSCRM3 predicts that at $J$=8 the rotation transitions to a uniform rotation about the $x$ axis. This transition (i.e., quenching of the wobbly rotation) reduces the energy-level spacing between $J$=6 and 8 as is observed in experiments [2,4]. This result resolves the long-standing mystery as to the cause of this energy-level spacing reduction, which is not predicted by CCRM3. It is shown that the residuals of the square of the angular momentum and two-body interaction, and to a lesser extent the spin-orbit interaction in MSCRM3 remove the remaining discrepancy between the predicted and measured excitation energies in $^{20}$Ne.

For triaxial $^{24}$Mg, MSCRM3 and CCRM3 predict band termination at $J$=4 and 8 (depending on initial values of the angular-velocity polar angles) when the nucleus becomes axially symmetric. However, MSCRM3 also predicts band termination at $J$=12 (observed in shell-model calculations [36] and experiments [37]) when $^{24}$Mg and its rotation become spherically symmetric about one of the principal axes.

This latter band termination is not predicted by CCRM3.

For some values of the angular-velocity polar angles at the start of the iteration process, MSCRM3 predicts no band termination in triaxial $^{24}$Mg whereas CCRM3 does, and vice versa.

In a number of cases for triaxial $^{24}$Mg, MSCRM3 predicts no rotation of any kind over a range of angular-velocity polar angles whereas CCRM3 predicts uniform rotation.

In a number of cases MSCRM3 predicts band termination whereas CCRM3 does not and vice versa.

Neither MSCRM3 nor CCRM3 predicts any physically meaningful rotations in triaxial $^{28}$Si for any values of $J$ and initial values of the angular-velocity polar angles.

No steady stable collective rotation of any kind is predicted by neither MSCRM3 nor CCRM3 for oblate $^{24}$Mg and $^{28}$Si.

The above results show that CCRM3 does not predict any of the three-dimensional phenomena predicted by MSCRM3 such as reduced energy level spacing in $^{20}$Ne and band terminations in prolate $^{28}$Si and triaxial $^{24}$Mg at $J$=12 when the nuclei and their rotations become spherically symmetric. These three-dimensional phenomena arise from the effects of the microscopic angular velocity in MSCRM3. We, therefore, conclude that CCRM3 is effectively a uniaxial rotation model. Therefore, using CCRM3 or its uniaxial version, one would miss capturing three-dimensional rotation phenomena and the associated excitation energies and quadrupole moments. Furthermore, MSCRM3 Hamiltonian contains residual terms that are normally negligeable but may become important for triaxial rotations at high spins and in back-bending regions where the angular velocity and momentum undergo large changes.

It must be emphasized that inclusion, in the models, of pairing and spin-orbit interactions and residuals of two-body interaction and the square of the angular momentum can change the above preliminary results predicted by MSCRM3 and CCRM3. Therefore, we are including these interactions in MSCRM3 in a planned calculation of nuclear rotational properties to achieve a physically meaningful comparison between the measured and predicted results. It may also be useful to generalize MSCRM3 to include vibrational excitations.

**Acknowledgements**

One of the authors (P.G.) would like to thank Professor M. A. Caprio at the University of Notre Dame as well as others for providing valuable comments on the manuscript.